\documentclass{ws-mpla}

\begin{document}

\title{Five-Dimensional Cosmological Scaling Solution }
\author{BAORONG CHANG, HONGYA LIU\thanks{
Corresponding author}, HUANYING LIU, LIXIN XU}

\maketitle

\pub{Received (15 June 2004)}{Revised (5 September 2004)}

\begin{abstract}
A five-dimensional Ricci-flat cosmological solution is studied by assuming
that the induced 4D matter contains two components: the usual fluid for dark
matter as well as baryons and a scalar field with an exponential potential
for dark energy. With use of the phase-plane analysis it is shown that there
exist two late-time attractors one of which corresponds to a universe
dominated by the scalar field alone and the other is a scaling solution in
which the energy density of the scalar field remains proportional to that of
the dark matter. It is furthermore shown that for this 5D scaling solution
the universe expands with the same rate as in the 4D FRW models and not
relies on which 4D hypersurface the universe is located in the 5D manifold.
\end{abstract}


\markboth{Baorong Chang, Hongya Liu, Huanying Liu and Lixin Xu}
{Five-Dimensional Cosmological Scaling Solution}

\catchline{}{}{}{}{} %
%%%%%%%%%%%%%%%%%%%%%%%%%%%%%%%%%%%%%%%%%%%%%%%%%%%%%%%%%%%%%%%%%%%%

\address{Department of Physics, Dalian University of Technology \\
Dalian, Liaoning, 116024, P. R. China \\
hyliu@dlut.edu.cn}

\ccode{PACS Nos.: 04.50.+h, 98.80.-k.}

\section{Introduction}

Scalar fields play a central role in modern cosmology in driving inflation
of the early universe and describing dark energy of the present epoch.$%
^{1,2} $ Observations predict that our universe contains roughly one-third
of dark matter and baryons and two- thirds of dark energy.$^{3}$ Within the
standard Friedmann-Robertson-Walker (FRW) models it was shown that there
exist scaling solutions that are the unique late-time attractors whenever
they exist.$^{4,5,6}$ Here, in this paper, we wish to look for scaling
solutions in higher-dimensional cosmological models.

In Kaluza-Klein theories as well as in brane world scenarios, our 4D
universe is believed to be embedded in a higher-dimensional manifold. One of
these models is the Ricci-flat 5D cosmological solutions presented by Liu
and Wesson.$^{7}$ This model is 5D Ricci-flat, implying that it is empty
viewed from 5D. However, as is known from the induced matter theory,$^{8,9}$
4D Einstein equations with matter could be recovered from 5D equations in
apparent vacuum. This approach is guaranteed by Campbell's theorem that any
solution of the Einstein equations in N-dimensions can be locally embedded
in a Ricci-flat manifold of (N+1)-dimensions.$^{10}$ In section II, we
suppose the 4D induced matter be composed of a perfect fluid and a scalar
field. In section III, we use phase-plane analysis to study the evolutions
of the model. In section IV we study the scaling solution. Section V is a
short discussion.

\section{5D Solution With An Induced Scalar Field}

An exact 5D cosmological solution was given firstly by Liu and Mashhoon in
Ref. 11 and restudied by Liu and Wesson in Ref. 7. This solution reads%
\begin{eqnarray}
dS^{2}&=& B^{2}(t,y)dt^{2}-A^{2}(t,y)\left[\frac{dr^{2}}{1-kr^{2}}%
+r^{2}(d\theta ^{2}+\sin ^{2}\theta d\phi ^{2})\right] -dy^{2},  \nonumber \\
B&=&\frac{1}{\mu}\frac{\partial A}{\partial t}\equiv \frac{\dot{A}}{\mu}
\label{5Dmatric} \\
A^{2}&=&(\mu ^{2}+k)y^{2}+2\nu y+\frac{\nu ^{2}+K}{\mu ^{2}+k},  \nonumber
\end{eqnarray}%
where $\dot{A}=(\partial /\partial t)A$, $\mu =\mu (t)$ and $\nu =\nu (t)$
are two integration functions of $t$ , $k$ is the 3-D curvature index $%
(k=\pm 1,0)$ and $K$ is a constant. This solution satisfies the 5D vacuum
equation $R_{AB}=0,$ with the three invariants being%
\begin{equation}
I_{1}\equiv R=0, I_{2}\equiv R^{AB}R_{AB}=0, I_{3}\equiv R^{ABCD}R_{ABCD}=%
\frac{72K^{2}}{A^{8}}.  \label{b}
\end{equation}%
So $K$ determines the curvature of the 5D manifold.

The 5D line element in (\ref{5Dmatric}) contains the 4D one,%
\begin{equation}
ds^{2}=g_{\alpha \beta }dx^{\alpha }dx^{\beta }=B^{2}dt^{2}-A^{2}\left[
\frac{dr^{2}}{1-kr^{2}}+r^{2}(d\theta ^{2}+\sin ^{2}\theta d\phi ^{2})\right]%
.  \label{c}
\end{equation}%
Using this 4D metric we can calculate the 4D Einstein tensor by $%
^{(4)}G_{\beta }^{\alpha }\equiv ^{(4)}\smallskip R_{\beta }^{\alpha
}-\delta _{\beta }^{\alpha }$ $^{(4)}R/2$. Its non-vanishing components are%
\begin{eqnarray}
&^{(4)}G_{0}^{0}=\frac{3(\mu ^{2}+k)}{A^{2}},&  \nonumber \\
&^{(4)}G_{1}^{1}=^{(4)}G_{2}^{2}=^{(4)}G_{3}^{3}=\frac{2\mu \dot{\mu}}{A\dot{%
A}}+\frac{(\mu ^{2}+k)}{A^{2}}.  \label{d}
\end{eqnarray}%
Generally speaking, the Einstein tensor in (\ref{d}) can give a 4D effective
or induced energy-momentum tensor $^{(4)}T_{\alpha }^{\beta }$ via $%
^{(4)}G_{\alpha }^{\beta }=\varkappa ^{2\,(4)}T_{\alpha }^{\beta }$ with $%
\varkappa ^{2}=8\pi G$.

In previous works,$^{7,12}$ this energy-momentum tensor was modelled by a
perfect fluid with density $\rho $ and pressure $p$, plus a variable
cosmological term $\Lambda $. This $\Lambda $ could be served to describe
dark energy. In this paper, we let the energy-momentum tensor consist of two
parts:
\begin{eqnarray}
&T_{\mu \nu }=T_{\mu \nu }^{m}+T_{\mu \nu }^{\phi },  \nonumber \\
&T_{\mu \nu }^{m}=(\rho _{m}+p_{m})u_{\mu }u_{\nu }-p_{m}g_{\mu \nu },&
\label{fluid} \\
&T_{\mu \nu }^{\phi }=\partial _{\mu }\phi \partial _{\nu }\phi -g_{\mu \nu
}[\frac{1}{2}g^{\alpha \beta }\partial _{\alpha }\phi \partial _{\beta }\phi
-V(\phi )],  \nonumber
\end{eqnarray}%
where $T_{\mu \nu }^{m}$ represents a perfect fluid and $T_{\mu \nu }^{\phi
} $ represents a scalar field. Then, from the 4D conservation laws $%
T_{\,\,\,\,\,\mu ;\nu }^{(m)\,\nu }=0$ and $T_{\,\,\,\,\mu ;\nu }^{(\phi
)\,\nu }=0$, we obtain%
\begin{eqnarray}
&\dot{\rho}_{m}+\frac{3\,\dot{A}}{A}(\rho _{m}+p_{m})=0,  \nonumber \\
&\ddot{\phi}+(\frac{3\,\dot{A}}{A}-\frac{\dot{B}}{B})\dot{\phi}+B^{2}\frac{dV%
}{d\phi }=0.  \label{f}
\end{eqnarray}%
\ The Hubble parameter $H$ and the deceleration parameter $q$ should be
defined in terms of the proper time as.$^{12}$
\begin{equation}
H\equiv \frac{\dot{A}}{AB}=\frac{\mu }{A}, q=-\frac{A\,\dot{\mu}}{\mu \,\dot{%
A}}.  \label{decelerate}
\end{equation}%
Using this we can recover the corresponding Friedmann equations as follows:%
\begin{eqnarray*}
&H^{2}+\frac{k}{A^{2}}=\frac{\varkappa ^{2}}{3}\left( \rho _{m}+\rho _{\phi
}\right) & \\
&\dot{H}=-\frac{\varkappa ^{2}}{2}B(\rho _{m}+p_{m}+\rho _{\phi }+p_{\phi
}),&
\end{eqnarray*}%
where%
\begin{equation}
\rho _{\phi }\equiv \frac{1}{2}\frac{\dot{\phi}^{2}}{B^{2}}+V(\phi ),
p_{\phi }\equiv \frac{1}{2}\frac{\dot{\phi}^{2}}{B^{2}}-V(\phi ).  \label{i}
\end{equation}

\section{Phase-plane Analysis Of The Solution}

From (\ref{5Dmatric}) we see that in general we have $B(t,y)\neq const$, so $%
t$ is not the proper time. This leads to the only difference between our
results (\ref{f})-(\ref{i}) and those of the standard FRW models. Similar as
in Ref. 5, we define $x$ and $y$ as

\begin{equation}
x\equiv \frac{\varkappa \,\dot{\phi}}{\sqrt{6}BH}, y\equiv \frac{\varkappa
\sqrt{V}}{\sqrt{3}H}.  \label{j}
\end{equation}%
For an exponential potential $V=V_{0}\exp (-\lambda \varkappa \phi )$ and a
spatially flat universe $(k=0)$, we find that the evolution equation for $x$
and $y$ are of the same form as in Ref. 5,
\begin{eqnarray}
x^{^{\prime }}&=&-3x+\lambda \sqrt{\frac{3}{2}}y^{2}+\frac{3}{2}%
x[2x^{2}+\gamma _{m}(1-x^{2}-y^{2})],  \nonumber \\
y^{^{\prime }}&=&-\lambda \sqrt{\frac{3}{2}}x y+\frac{3}{2}y[2x^{2}+\gamma
_{m}(1-x^{2}-y^{2})],  \label{xandy}
\end{eqnarray}%
where a prime denotes a derivative with respect to the logarithm of the
scale factor, $N=\ln A$. Define
\begin{equation}
\Omega _{m}=\frac{\varkappa ^{2}\rho _{m}}{3H^{2}}, \Omega _{\phi }=\frac{%
\varkappa ^{2}\rho _{\phi }}{3H^{2}},  \label{l}
\end{equation}%
then (\ref{xandy}) gives $\Omega _{m}+\Omega _{\phi }=1$ and (\ref{j}) gives
$x^{2}+y^{2}=\Omega _{\phi }$. Meanwhile, from (\ref{i}), the effective
equation of state for the scalar field gives%
\begin{equation}
\gamma _{\phi }\equiv \frac{\rho _{\phi }+p_{\phi }}{\rho _{\phi }}=\frac{%
2x^{2}}{x^{2}+y^{2}}.  \label{m}
\end{equation}

From the analysis of Ref. 5 we know that there are five fixed points
(critical points) corresponding to $x^{^{\prime }}=0$ and $y^{^{\prime }}=0$
in the plane-autonomous system, including two stable nodes.$^{5}$ The two
stable nodes represent two possible late-time attractor solutions and,
therefore, are of particular physical interest. We list them in the
following.

(1). Late-time attractor solution dominated by the scalar field alone.

\ \ This solution corresponds to the stable node%
\begin{equation}
x=\lambda /\sqrt{6}, y=\sqrt{1-\lambda ^{2}/6}, with \lambda ^{2}<6
\label{n}
\end{equation}%
for which we have $x^{2}+y^{2}=1$, then we get%
\begin{equation}
\Omega _{m}=0, \Omega _{\phi }=1,\gamma _{\phi }=\lambda ^{2}/3.  \label{o}
\end{equation}%
So the universe is dominated by the scalar field alone and expands with a
power-law. It can be shown that if, furthermore, $\lambda ^{2}<2$, the
universe is accelerating.

(2). Scaling solution.

\ \ It corresponds to the stable node%
\begin{equation}
x=\sqrt{\frac{3}{2}}\frac{\gamma _{m}}{\lambda }, y=\sqrt{\frac{3(2-\gamma
_{m})\gamma _{m}}{2\lambda ^{2}}}, with \lambda ^{2}>3\gamma _{m}  \label{p}
\end{equation}%
for which we have%
\begin{equation}
\Omega _{m}=\frac{\lambda ^{2}-3\gamma _{m}}{\lambda ^{2}}, \Omega _{\phi }=%
\frac{3\gamma _{m}}{\lambda ^{2}}, \gamma _{\phi }=\gamma _{m}.  \label{q}
\end{equation}%
So this represents a scaling solution where the dark energy density of the
scalar field is proportional to that of the baryons and dark matter.

\section{5D Scaling Solution}

It is known that in the standard spatially-flat FRW models the scaling
solution (\ref{p}) corresponds to a power-law expansion. Here let us check
how the universe expands in the 5D model (\ref{5Dmatric}).

Firstly, by substituting the value of $x$ of (\ref{p}) into its definition
in (\ref{j})\ and using (\ref{decelerate}), we find that the scalar field $%
\phi $ can be integrated, giving%
\begin{equation}
\phi =\frac{3\gamma }{\kappa \lambda }\ln A+C_{1},  \label{r}
\end{equation}%
Thus we obtain the exponential potential as%
\begin{equation}
V\equiv V_{0}\exp (-\lambda \varkappa \phi )=C_{2}V_{0}A^{-3\gamma _{m}}
\label{v}
\end{equation}%
Meanwhile, substituting the value of $y$ of (\ref{p}) into its definition in
(\ref{j}), and then using (\ref{v}), we get a constraint between $A$ and $%
\mu $ as%
\begin{equation}
\mu ^{2}=C_{3}^{2}A^{2-3\gamma _{m}}.  \label{s}
\end{equation}

Consider the 5D metric (\ref{5Dmatric}). The form $Bdt=(\dot{A}/\mu )dt$ is
invariant under an arbitrary coordinate transformation $t\rightarrow \tilde{t%
}(t)$. This would enable us to choose the coordinate $t$ such that although $%
t$ does not equal to the proper time $\tau $ globally, it can tend to the
proper time as $t\rightarrow \infty $. Be aware that in the 4D standard FRW
models the scaling solution (\ref{p}) corresponds to the expansion $%
a(t)\propto t^{2/(3\gamma _{m})}$ and $H=2/(3\gamma _{m}t)$. Thus we obtain
the late-time approximation of the 5D scaling solution (\ref{p}) as:%
\begin{equation}
A\approx t^{\frac{2}{3\gamma _{m}}}, B\approx 1, \mu \approx \frac{2}{%
3\gamma _{m}}t^{\frac{2}{3\gamma _{m}}-1}, H=\frac{\mu }{A}\approx \frac{2}{%
3\gamma _{m}t}.  \label{t}
\end{equation}%
For $k=0$, the 5D solution (\ref{5Dmatric}) reads%
\begin{equation}
\ A^{2}=\mu ^{2}y^{2}+2\nu y+\frac{\nu ^{2}+K}{\mu ^{2}}.  \label{u}
\end{equation}%
Using (\ref{t}) in (\ref{u}), we find%
\begin{equation}
\nu \approx \frac{2}{3\gamma _{m}}t^{\frac{4}{3\gamma _{m}}-1}.  \label{w}
\end{equation}

\section{Discussion}

In this paper we have used the phase-plane analysis to study the stability
of evolution of a 5D cosmological model. From Eqs.(\ref{t})-(\ref{w}) we can
draw three conclusions:

(1). The scaling solution (\ref{p}), as a late-time attractor of the
evolution of the universe, is the same in both the 4D FRW models and the 5D
induced matter theory.

(2). In the 5D model (\ref{t})-(\ref{w}), the expansion rate $A$ and $H$ of (%
\ref{t}) seems not rely on the value of $y$. In other words, our 4D
universe, no matter it is situated on which hypersurface of $y$ in the 5D
manifold, looks similar and expands with almost the same rate, in late times.

(3). The constant $K$ in (\ref{u}) represents the curvature of the 5D
manifold (see Eq.(\ref{b})). It is reasonable to believe that $K=0,\pm 1$
may correspond to three different topologies for the 5D manifold (\ref%
{5Dmatric}). However, from (\ref{t})-(\ref{w}), we find that $K$ does not
affect $A$ and $H$ so much in late-times.

(4). In Eq. (\ref{t}), if $\gamma _{m}=1/3$, we have $B\approx 1$ and $%
A\approx t^{2}$, and our model approaches to the Milne model which has no
event or particle horizon, and so it is of particular interest.

It was shown$^{7,12}$ that in the early-time of the universe the 5D solution
(\ref{5Dmatric}) deviates from the standard FRW models greatly. For instance
it may have a big bounce rather than a big bang and before the bounce the
universe is contracting. Meanwhile, the value of the 5D curvature $K$ is
sensitive for having or not having a bounce. However, in the late time, we
have $B\longrightarrow 1$ and the coordinate time $t$ gives back to the
cosmic time $\tau $,\ so the bounce model approaches to the FRW model. As a
late-time attractor, the scaling solution seems not show up noticeable
deviations from the standard FRW models. Further study along this line is
needed to distinguish the two theories.

\section{Acknowledgments}

This work was supported by National Natural Science Foundation (10273004)
and National Basic Research Program (2003CB716300) of P. R. China.


\section{References}

\begin{thebibliography}{99}
\bibitem{1} A. H. Guth, \textit{Phys. Rev.} \textbf{D23,} 347 (1981); A. D.
Linde, \textit{Phys. Lett.} \textbf{B108,} 389 (1982); A. D. Linde, \textit{%
Phys. Lett.} \textbf{B129,} 77 (1983).

\bibitem{2} K. Coble, S. Dodelson and J. A. Frieman, \textit{Phys. Rev.}
\textbf{D55,} 1851 (1997), astro-ph/9608122; R. R. Caldwell and P. J.
Steinhardt, \textit{Phys. Rev.} \textbf{D57,} 6057 (1998), astro-ph/9710062;
I. Zlatev, L. M. Wang and P. J. Steinhardt, \textit{Phys. Rev.} \textit{Lett.%
} \textbf{82,} 896 (1999), astro-ph/9807002; P. J. Steinhardt, L. M. Wang
and I. Zlatev, \textit{Phys. Rev.} \textbf{D59,} 123504 (1999),
astro-ph/9812313.

\bibitem{3} W. L. Freedman, M. S. Turner, \textit{Rev. Mod. Phys.} \textbf{75%
}, 1433 (2003), astro-ph/0308418.

\bibitem{4} A. R. Liddle, R. J. Scherrer, \textit{Phys. Rev.} \textbf{D59,}
023509-1 (1998), astro-ph/9809272.

\bibitem{5} E. J. Copeland, A. R. Liddle, and D. Wands, \textit{Phys. Rev.}
\textbf{D57}, 4686 (1998), gr-qc/9711068.

\bibitem{6} R. J. van den Hoogen, A. A. Coley and D. Wands, \textit{Class.
Quant. Grav.} \textbf{16, }1843 (1999), gr-qc/9901014; I. P. C. Heard and D.
Wands, \textit{Class. Quant. Grav.} \textbf{19, } 5435 (2002),
gr-qc/0206085; Z. K. Guo, Y. S. Piao, R. G. Cai and Y. Z. Zhang, \textit{%
Phys. Lett.} \textbf{B576, }12 (2003), hep-th/0306245; Z. K. Guo, Y. S. Piao
and Y. Z. Zhang, \textit{Phys. Lett.} \textbf{B568, }1 (2003),
hep-th/0304048.

\bibitem{7} H. Y. Liu and P. S. Wesson, \textit{Astrophys. J.} \textbf{562,}
1 (2001), gr-qc/0107093.

\bibitem{8} J. M. Overduin and P. S. Wesson, \textit{Phys. Rep.} \textbf{283}%
, 303 (1997), gr-qc/9805018.

\bibitem{9} P. S. Wesson, \textit{Space-Time-Matter} (World Scientific,
1999).

\bibitem{10} J. E. Campbell, \textit{A Course of Differential Geometry}
(Clarendon, 1926).

\bibitem{11} H. Y. Liu and B. Mashhoon, \textit{Ann. Phys.} \textbf{4}, 565
(1995).

\bibitem{12} H. Y. Liu, \textit{Phys. Lett.} \textbf{B560, }149 (2003),
hep-th/0206198; B. L. Wang, H. Y. Liu and L. X. Xu, \textit{Mod. Phys. Lett.}
\textbf{A19, }449 (2004), (gr-qc/0304093); X. L. Xin, H. Y. Liu and B. L.
Wang, \textit{Chin. Phys. Lett.} \textbf{20}, 995 (2003), (gr-qc/0304049).
\end{thebibliography}
\end{document}